\documentclass[aps,nofootinbib,preprintnumbers,twocolumn,superscriptaddress,prd]{revtex4-1}

\usepackage{amsmath}
\usepackage{amssymb}
\usepackage{slashed}
\usepackage{epsfig}
\usepackage{bbm,bm}
\usepackage{hyperref}
\usepackage{color}
\usepackage{comment}
\usepackage{amsfonts}
\usepackage{dsfont}

\newcommand{\be}{\begin{eqnarray}}
\newcommand{\ee}{\end{eqnarray}}
\newcommand{\Eqref}[1]{Eq.~\eqref{#1}}

\begin{document}

\title{Lee-Yang model from the functional renormalization group}

%\date{\today}

\author{Luca Zambelli}
\email{luca.zambelli@uni-jena.de}
\affiliation{\mbox{\it Theoretisch-Physikalisches Institut, Friedrich-Schiller-Universit{\"a}t Jena,}
\mbox{\it D-07743 Jena, Germany}}

\author{Omar Zanusso}
\email{omar.zanusso@uni-jena.de}
\affiliation{\mbox{\it Theoretisch-Physikalisches Institut, Friedrich-Schiller-Universit{\"a}t Jena,}
\mbox{\it D-07743 Jena, Germany}}

%\pacs{}

%%%%%%%%%%%%%%%%%%%%%%%%%%%%%%%%%%%%%%%%

\begin{abstract} 
We investigate the critical properties of the Lee-Yang model in less than six spacetime dimensions using truncations of the functional renormalization group flow.
We give estimates for the critical exponents, study the dependence on the regularization scheme,
and show the convergence of our results for increasing size of the truncations in four and five dimensions.
While with our truncations it is numerically challenging to approach the three-dimensional case,
we provide a simple approximation which allows us to qualitatively study the Lee-Yang model in two and three dimensions,
and use it to argue the existence of further nonunitary multicritical theories including one which is relevant for the universality class of the Blume-Capel model.
\end{abstract}

\maketitle

%%%%%%%%%%%%%%%%%%%%%%%%%%%%%%%%%%%%%%%%
\section{Introduction}
\label{sec:intro}
%%%%%%%%%%%%%%%%%%%%%%%%%%%%%%%%%%%%%%%%

The cubic scalar field theory with
classical Lagrangian
${\cal L}=(\partial\varphi)^2+\lambda\varphi^3$, despite its simplicity, shows a surprisingly rich spectrum of phenomena.
Historically, this model in $d=6$ spacetime dimensions has served as a prototype of an asymptotically free theory~\cite{Macfarlane:1974vp}
and was used as a toy model to illustrate several aspects of QCD \cite{Cardy:1974af,Ma:1975vn,Cornwall:1995dr,Mikhailov:1997zg}.
For instance, it is remarkable
that a particular cubic theory of ${\rm N}$ scalars, which can be mapped to the $({\rm N}+1)$-states
Potts model \cite{Zia:1975ha}, in the limit ${\rm N}\to 0$ belongs to the same universality class of the Reggeon field theory and the directed percolation model \cite{Cardy:1980rr}. 

Indeed, the most fertile area of application for this kind of theories is offered by critical phenomena.
A celebrated case of study is that of a single real scalar field, known as the Lee-Yang model,
which has upper critical dimension $d=6$ and provides a Ginzburg-Landau description of a peculiar critical point \cite{Fisher:1978pf},
the Yang-Lee edge singularity~\cite{Kortman:1971zz}.
This model provides an outstanding example of universality since, while stemming from the thermodynamic properties of 
Ising ferromagnets in $d$ dimensions~\cite{Yang:1952be},
it governs the negative-activity singularity of fluid models with repulsive-core interactions in the same dimensionality~\cite{Lai:1995,Park:1999},
directed branched polymers (directed loop-free lattice animals) in $(d+1)$ dimensions~\cite{Cardy:1982},
Anderson localization phenomena~\cite{Lubensky:1981}
and isotropic branched polymers (undirected lattice animals)~\cite{Parisi:1980ia}, both in $(d+2)$ dimensions.
The latter correspondence descends from a beautiful connection with a 
${\cal{N}}=1$ supersymmetric model
in $(d+2)$ dimensions~\cite{Parisi:1979ka}.

Away from criticality the simple cubic model seems to be plagued by a terminal
illness: the instability of the bare potential which is believed to generically result 
into an unbounded-from-below energy spectrum.
However, this conclusion has been questioned by several authors, and
two notable mechanisms to stabilize the theory have been proposed.
On the one hand, the stability of the model might be tied to ${\cal PT}$ symmetry~\cite{Bender:2004sa,Bender:2012ea}, which
is present at criticality, as signaled, for instance, by the imaginary value 
of the cubic coupling $\lambda$.
On the other hand, the theory might be rescued through the inclusion of additional degrees of freedom, such as N scalars
with $O({\rm N})$ symmetry.

This latter possibility has received some attention in the past two years:
on the basis of the AdS/CFT duality,
it has been suggested that there exist $O({\rm N})$-invariant conformal models which are unitary in $d<6$ for N big enough~\cite{Fei:2014yja,Giombi:2014xxa,Diab:2016spb}.
These models have been perturbatively analyzed in $d=(6-\epsilon)$
dimensions by means of a parametrization in terms of an additional auxiliary $O({\rm N})$
scalar singlet with \emph{cubic} interaction~\cite{Fei:2014yja,Giombi:2014xxa,Diab:2016spb}. 
The cubic sector generates a nontrivial fixed point in $d<6$, which is real for large values of N
and which thus offers a cure to the triviality problem in $4<d<6$.
The issue of the stability of these models remains still open within perturbation theory,
and nonperturbative methods have already been applied to try to settle it~\cite{Percacci:2014tfa,Mati:2014xma,Eichhorn:2016hdi}.
From our perspective, the nonperturbative study of the Lee-Yang model is a complementary 
and more elementary task that should be addressed.
Together, all these lines of research complement our general understanding of scalar field theories,
which are at the heart of high-energy physics through the Higgs sector of the standard model,
and of modern theoretical physics through the AdS/CFT correspondence.

Similarly to the quartic model, the cubic scalar model is  generally not exactly solvable.
The main properties of the critical theory are exactly known
only in $d=1$~\cite{Fisher:1978pf} and $d=2$~\cite{Cardy:1985yy}.
More precisely, in the latter case it has been argued that the two-dimensional conformal field theory
corresponding to the Lee-Yang universality class is the minimal ${\cal M}(2,5)$ model
which arises as a representation of the Virasoro algebra \cite{Belavin:1984vu}.
Interestingly, it has been argued that the Lee-Yang model is only the first element of
a sequence of two-dimensional multicritical theories \cite{vonGehlen:1994rp}
which are captured by the minimal models ${\cal M}(2,2n+3)$ for $n\in \mathbb{N}_0$
and of which the critical phases are reached by tuning the magnetic field to a purely imaginary value \cite{Belavin:2003pu}.
The second element of this sequence ${\cal M}(2,7)$ describes the tricritical phase of the Blume-Capel model for spin chains
and thus could have application to models of atomic mixtures \cite{BC}.

In more than two dimensions, the full understanding of the dynamics of the Lee-Yang theory requires an intricate web of different approximations.
Several methods have been applied to the study of the critical model,
ranging from the $(6-\epsilon)$ expansion~\cite{deAlcantaraBonfim:1980pe,Gracey:2015tta} to strong coupling expansions~\cite{Butera:2012tq},
and from lattice simulations~\cite{Lai:1995,Park:1999,Hsu:2005} to the conformal 
bootstrap~\cite{Gliozzi:2014jsa,Pismenskii:2015xxg,Nakayama:2016cim,Hasegawa:2016piv,Gopakumar:2016cpb}.
As a complementary approach, the use of functional renormalization group (FRG) techniques
is very well suited to this problem, since it can be naturally generalized to any continuous dimension (from high to low $d$) 
without the need of resummations or extrapolations, in any regime (both weakly or strongly coupled, massive or massless),
and to a wide family of models at once including arbitrary real or complex couplings or bare actions (e.g.\ generic interacting potentials)~\cite{ReviewRG}.

A first exploratory FRG analysis of the Yang-Lee edge singularity
has recently appeared~\cite{An:2016lni}. The results of this investigation pave the way for
a more extensive and in-depth analysis of the
universal properties of this model, which should be tested against the change of both approximation strategies and regulator choices,
as well as improvements of the numerical analysis. The present work goes precisely in this direction.

The plan of this paper is as follows.
In Sec.~\ref{sec:FRG}, we describe the FRG method, the regularization scheme, 
and the approximations within which we perform our analysis.
In Sec.~\ref{sec:derivative_expansion}, we obtain the most accurate FRG
estimates for the critical theory in $4\leq d < 6$ as of now, and report on issues that
future works will have to face in order to push this description to lower dimensions.
In Sec.~\ref{sec:imaginary_approximation}, we describe a simpler approximation that
captures the essence of the model for all dimensionalities $2\leq d < 6$
and shows a tower of multicritical $i\varphi^{2n+1}$ theories arising below their upper critical dimension.
In Sec.~\ref{sec:conclusions}, we summarize the results and outline the
future directions that should be taken.

%%%%%%%%%%%%%%%%%%%%%%%%%%%%%%%%%%%%%%%%
\section{Functional renormalization group: approximations and scheme}
\label{sec:FRG}
%%%%%%%%%%%%%%%%%%%%%%%%%%%%%%%%%%%%%%%%

To investigate the renormalization group (RG) flow of the cubic scalar
theory we will focus on truncations of the RG equation for an effective average action $\Gamma_k[\varphi]$
that flows according to RG time $t=\log{k}$.
The action $\Gamma_k[\varphi]$ is constructed by modifiying a standard path integral
through the inclusion of a masslike regulator with
kernel $R_k(x,y)$ in the field's propagator.
The RG scale $k$ separates the excitations of the field $\varphi$
into IR models, the propagation of which is suppressed by the cutoff, and UV modes, which are integrated out in the path integral.
The resulting functional generator of 1PI vertices $\Gamma_k[\varphi]$ is thus an effective
description of the physics of the UV modes
and interpolates with the standard effective action in the limit $k\to 0$, at which all modes are integrated out.

The regulator kernel $R_k(x,y)$ additively deforms the inverse propagator
\begin{equation}
\Gamma_k^{(2)}(x,y)=\frac{\delta^2\Gamma_k}{\delta\varphi(x)\delta\varphi(y)}.
\end{equation}
and allows one to write a RG equation that is exact, but has the simple structure of a one-loop equation~\cite{Wetterich:1992yh,Morris:1993qb,Bonini:1992vh,Ellwanger:1993mw,Nicoll:1977hi}
\begin{equation}\label{eq:Wetterich}
\partial_t\Gamma_k[\varphi]=\frac{1}{2}{\rm Tr}\left[\partial_t R_k\left(\Gamma_k^{(2)}[\varphi]+R_k\right)^{-1}\right].
\end{equation}
Here $\partial_t$ is a derivative at fixed $\varphi$, while products, inverses and traces (${\rm Tr}$) on the right hand side
stand for matrixlike operations, where continuous spacetime points $(x,y)$ replace discrete row/column labels.
\Eqref{eq:Wetterich} is usually not exactly solvable,
but it still allows for perturbative approximations,
both within the framework of the $\epsilon$ expansion~\cite{Wegner:1972ih,Nicoll:1974zza,Riedel:1986re,Rosten:2010vm,ODwyer:2007brp}
and at the upper critical 
dimension~\cite{Papenbrock:1994kf,Bonini:1996bk,Bonanno:1997dj,Pernici:1998tp,Kopietz:2000bh,Arnone:2003pa,Rosten:2010vm,Codello:2013bra}.
A different and complementary approximation strategy is based on nonperturbative truncations, which require projecting
the RG flow on a treatable subset of functionals, ideally as large as possible.

In this work, we will restrict ourselves to the parametrization
of the dynamics
\begin{equation}\label{eq:derivative_expansion}
\Gamma_k[\varphi]=\int\! d^dx \left[ \frac{1}{2} Z_k(\varphi)\partial_\mu\varphi\partial^\mu\varphi+V_k(\varphi)\right]
\end{equation}
which can be considered as the $O(\partial^2)$ of a systematic derivative
expansion of the effective action.
We will adopt as a regulator a step-linear shape function
that in (Euclidean) momentum space reads
\begin{equation}\label{eq:cutoff_p2}
R_k(p^2)=a\left(k^2-p^2\right)\theta\left(k^2-p^2\right)\,,
\end{equation}
where $\theta$ is the standard Heaviside function and $a>0$ is a free parameter.
This functional kernel is chosen to optimize the results of the truncation of~\Eqref{eq:derivative_expansion}
in scalar models like the present one~\cite{Litim:2001up}.

Our focus is on the nontrivial scaling solution with upper critical
dimension $d=6$, which corresponds to a non-Gau\ss ian fixed point (FP)
of the RG flow of dimensionless renormalized quantities
\begin{equation}
\begin{split}
& \overline{\varphi}_{\rm R}=k^{-\frac{d-2+\eta}{2}}\varphi\,,
\\
&
 z_k(\overline{\varphi}_{\rm R})=k^{\eta} Z_k(\varphi)\,,\qquad
 v_k(\overline{\varphi}_{\rm R})=k^{-d} V_k(\varphi)\,,
\end{split}
\end{equation}
where $\eta$ denotes the anomalous dimension of the field at the FP.
We choose the normalization $z_k(0)=1$, and probe the spurious
dependence of universal quantities on the overall normalization of the action,
induced by truncations of the theory space, by varying the regulator
parameter $a$.
Since we will always refer to dimensionless renormalized quantities at the floating scale $k$,
we will drop both the $k$ and ${\rm R}$ subscripts in what follows.

\begin{figure}[!htb] 
	\begin{center}
		\includegraphics[width=0.48\textwidth]{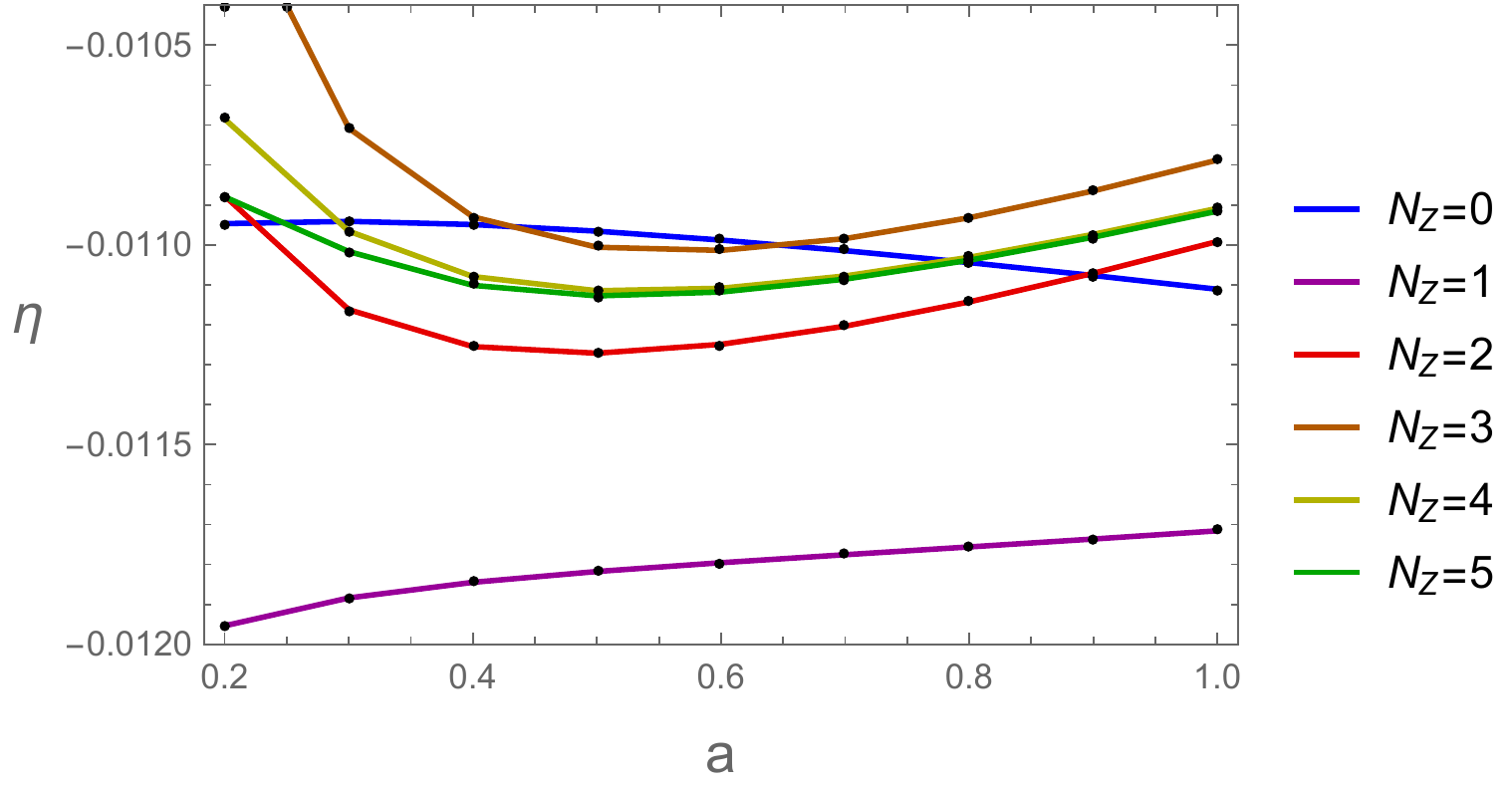}
		\caption{The critical anomalous dimension $\eta$ in $d=5.9$, as a function of the regulator parameter $a$,
			for several orders of the polynomial truncation $N_Z$ ($N_V=N_Z+3$).}
		\label{fig:eta_vs_a_59D}
	\end{center}
\end{figure}

The beta functionals for the truncation \eqref{eq:derivative_expansion} have been presented 
in~\cite{Morris:1994jc} using a power-law cutoff instead of \eqref{eq:cutoff_p2}.
In brief, the flow of the scale-dependent potential is computed from the constant-field limit of \eqref{eq:Wetterich},
while the flow of the wave function is computed from the flow of the two-point function, which is obtained by taking two functional derivatives of \eqref{eq:Wetterich}.
Our beta functionals are considerably more complex than those that appeared in \cite{Morris:1994jc},
both because of the step function and because of the presence of the additional parameter $a$.
The explicit form of these beta functionals is not particularly illuminating, so we will not provide it in the paper.
% we will provide them in the form of ancillary files.

%%%%%%%%%%%%%%%%%%%%%%%%%%%%%%%%%%%%%%%%
\section{Derivative expansion at $O(\partial^2)$}
\label{sec:derivative_expansion}
%%%%%%%%%%%%%%%%%%%%%%%%%%%%%%%%%%%%%%%%

In this section we analyze the coupled system of equations for $v(\phi)$ and $z(\phi)$.
We project the flow of these functions on the following polynomial basis
\begin{equation}\label{eq:polynomial_truncation}
	v(\varphi)=\sum_{n=1}^{N_V}\frac{\lambda_n}{n!}\varphi^n ,\quad \	z(\varphi)=\sum_{n=0}^{N_Z}\frac{z_n}{n!}\varphi^n .
\end{equation} 
The flow obeys the following properties \cite{Morris:1994jc}:
the equation for $\partial_t v(\varphi)$ depends on $v(\varphi)$, $v'(\varphi)$, $v''(\varphi)$ and $z(\varphi)$,
while the equation for $\partial_t z(\varphi)$ depends on $z''(\varphi)$, $v'''(\varphi)$, $z(\varphi)$, $z'(\varphi)$ and $z''(\varphi)$.
This translates into a precise hierarchy for the polynomial couplings.
The flow equation for $\lambda_i$  depends on couplings up to
$\lambda_{i+2}$ and $z_i$, while that for
$z_j$ involves parameters up to $\lambda_{j+3}$ and $z_{j+2}$.
However, $\lambda_1$ does not appear in the beta functions 
of all the other couplings, as it is immediately visible from the right hand side of~\Eqref{eq:Wetterich}.
%Hence, we can drop the beta-function of this linear term while searching for FPs.
Since the FP condition is a set of two second-order
ordinary differential equations, the solution
is completely determined by four numbers,
which can be chosen as 
$\{\lambda_2, \lambda_3, \eta, z_1 \}$.

Let $\{g_i\}=\{\lambda_1,\dots, \lambda_{N_V},z_0,\dots , z_{N_Z}\}$ be the set of all couplings
of our truncation and $\beta_i=\partial_t g_i$ be their beta functions.
The critical exponents $\theta_i$ can be computed by diagonalizing 
the stability matrix $M_{ij}$ at the FP
\begin{eqnarray}
\begin{split}
M_{ij}=\frac{\partial\beta_i}{\partial g_j}
\end{split}
\end{eqnarray}
and coincide with the negative of its eigenvalues.
\begin{figure}[!htb] 
	\begin{center}
		\includegraphics[width=0.48\textwidth]{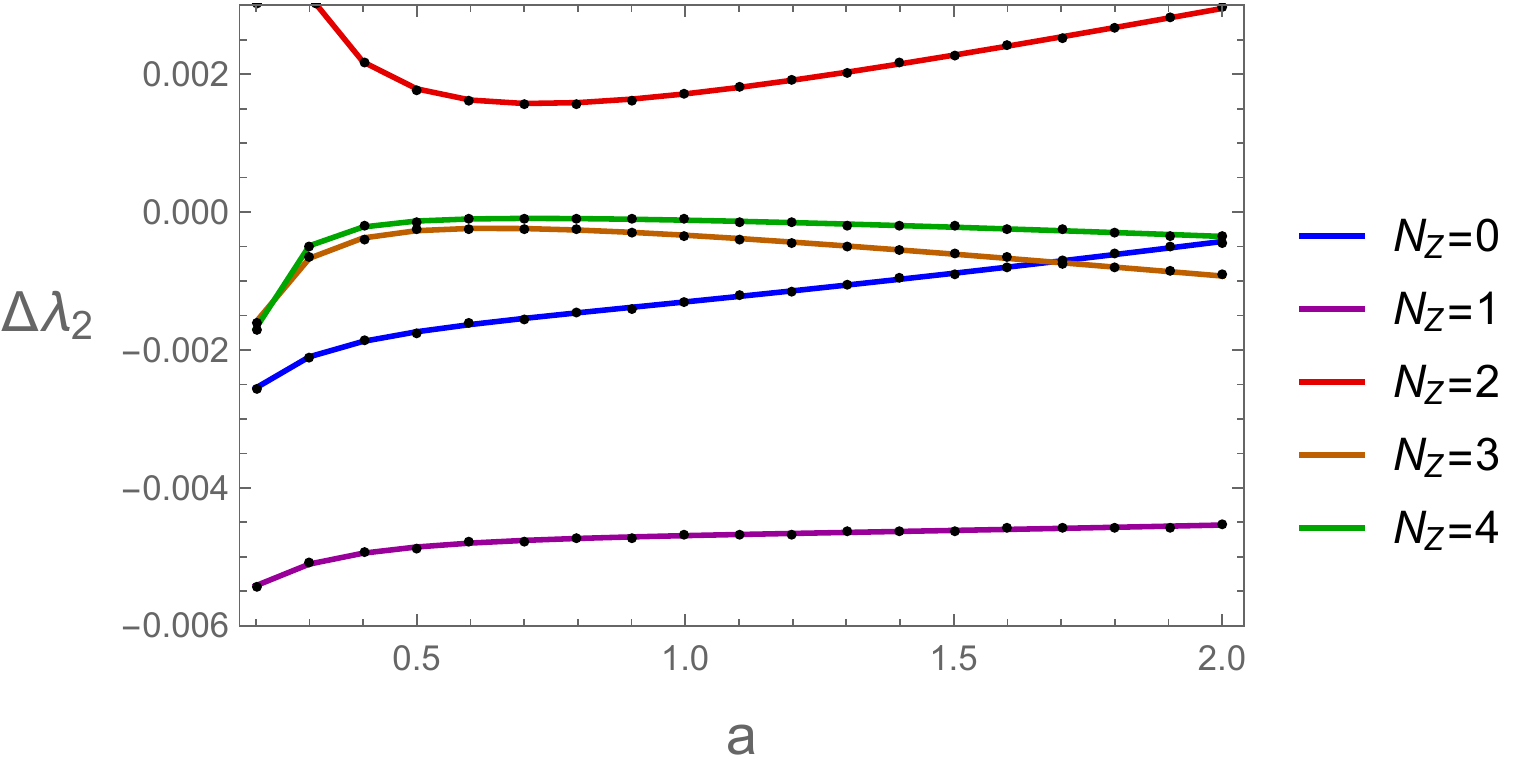}
		\caption{The violation of the hyperscaling relation for $\theta_2$, as defined in~\Eqref{eq:delta_lambda2},
			at the $d=5.9$ fixed point, as a 
			function of the regulator parameter $a$, for several orders of the polynomial truncation $N_Z$ ($N_V=N_Z+3$).}
		\label{fig:deltalambda2_vs_a_59D}
	\end{center}
\end{figure}
For each critical exponent, we can find a corresponding deformation,
\begin{eqnarray}
\begin{split}
v(\varphi)&=v_*(\varphi)+ \delta v_n(\varphi)\, {\rm e}^{-t\theta_n},\\
z(\varphi)&=z_*(\varphi)+ \delta z_n(\varphi)\, {\rm e}^{-t\theta_n},
\end{split}
\end{eqnarray}
which solves the linearized flow at the FP, the solutions of which are denoted with an asterisk.

% This is the polynomially truncated version of a functional
% eigenvalue problem for the linearized flow around the FP,
% which is obtained by inserting 
% %
% \begin{eqnarray}
% v(\varphi)&=&v_*(\varphi)+\epsilon\, \delta v_n(\varphi),\\
% z(\varphi)&=&z_*(\varphi)+\epsilon\, \delta z_n(\varphi),
% \end{eqnarray} 
% %
% in the flow equations, by linearizing in $\epsilon$,
% and demanding
% %
% \begin{eqnarray}
% \partial_t \delta v_n(\varphi) &=& -\theta_n \delta v_n(\varphi),\\
% \partial_t \delta z_n(\varphi) &=& -\theta_n \delta z_n(\varphi).
% \end{eqnarray} 
%

It is possible to prove very generally that the flow equations admit two relevant critical exponents,
which are 
related to the anomalous dimension through two scaling relations:
\begin{eqnarray}
\label{eq:scaling_rel_1}
 \theta_1
 =\frac{1}{\nu_c}=\frac{d+2-\eta}{2}\,,
 &\,\,\,&\delta v_1=\varphi,
 \,\,\,\,\,\,\,\,\,\,\delta z_1=0 \,,
\\
 \theta_2\,
 =\,\frac{1}{\nu}\,=\frac{d-2+\eta}{2}\,,
 &\,\,\,&\delta v_2=v'(\varphi),
 \,\delta z_2=z'(\varphi).
\label{eq:scaling_rel_2}
\end{eqnarray}
For example, to prove that \Eqref{eq:scaling_rel_2} holds, it suffices to compare the 
linearized equations to the $\varphi$ derivative of the FP 
equations~\cite{Osborn:2009vs,Hellwig:2015woa}.
While \Eqref{eq:scaling_rel_1} is preserved by the polynomial truncations 
of~\Eqref{eq:polynomial_truncation}, \Eqref{eq:scaling_rel_2}
is not.\footnote{A similar violation of an exact scaling relation by truncations of the FRG was
observed in~\cite{Eichhorn:2013zza} when studying Aharony's formula for the dimension 
of a certain operator at the decoupled FP of the $O({\rm N})\oplus O({\rm M})$ model.
In that case the violation could be removed, at least when considering $N_Z=0$ truncations,
by dropping the beta function of $z_0$ inside the stability matrix~\cite{Boettcher:2015pja}.
The same does not happen in the present work.}
In fact, one can use its violation
\begin{equation}\label{eq:delta_lambda2}
\Delta\theta_2=\frac{d-2+\eta}{2}-\theta_2
\end{equation}
as a measure of the quality of the truncation \cite{An:2016lni}.
Equations~\eqref{eq:scaling_rel_1} and~\eqref{eq:scaling_rel_2} also show that the anomalous dimension completely determines
all the relevant critical properties of the Lee-Yang model.\footnote{The implicit comparison is the Ising model which is characterized by $\eta$ and $\nu$, the latter one governing the scaling of the correlation length.} 
\begin{figure}[!htb] 
	\begin{center}
		\includegraphics[width=0.238\textwidth]{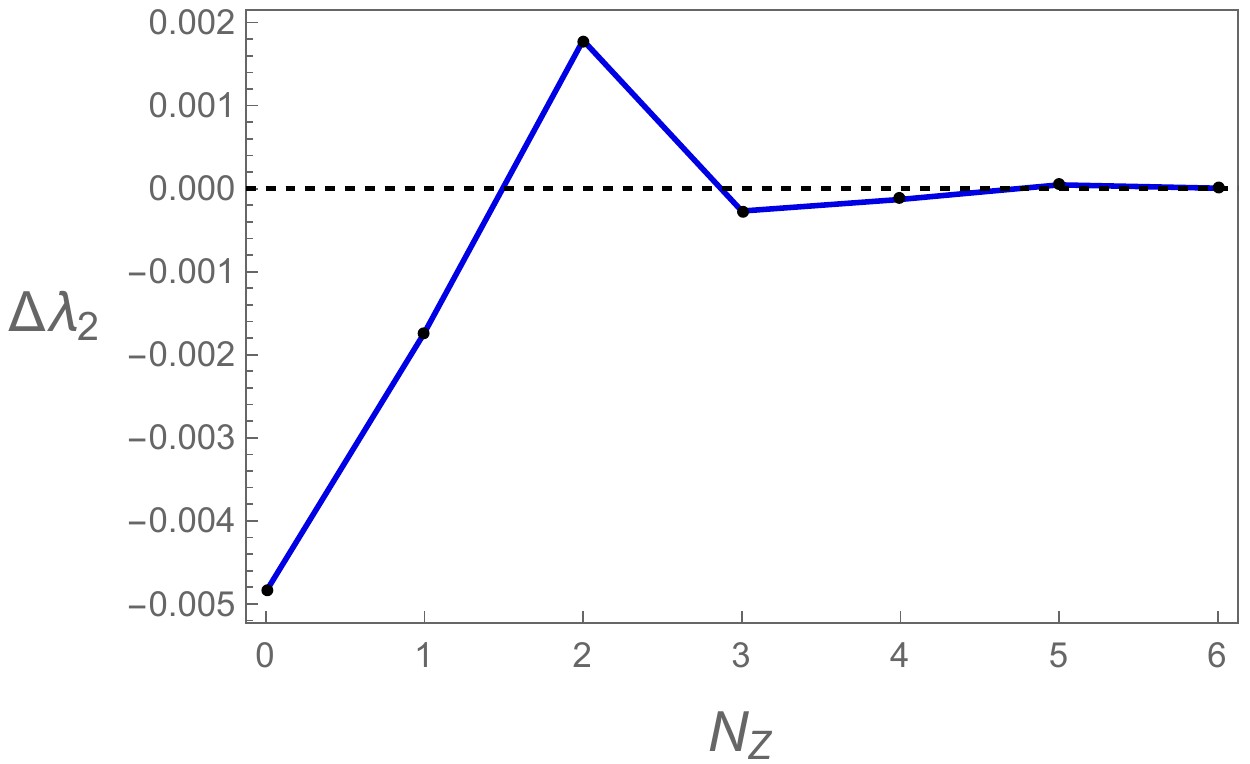}\hskip.5mm
		\includegraphics[width=0.238\textwidth]{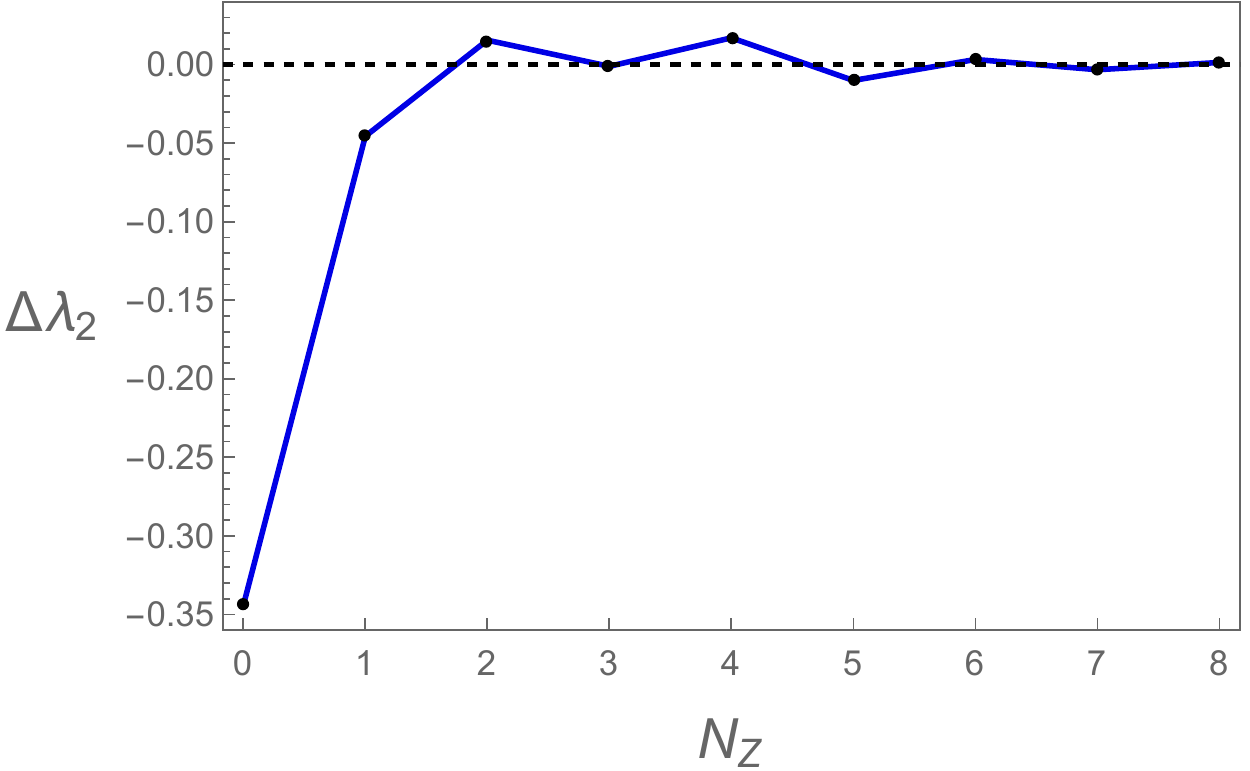}
		\caption{The violation of the hyperscaling relation for $\theta_2$, as defined in~\Eqref{eq:delta_lambda2},
			as a function of the order of the polynomial truncation, for $a=0.5$.
			Left panel: the $d=5.9$ fixed point.
			Right panel: the $d=5$ fixed point.}
		\label{fig:deltalambda2_vs_NZ}
	\end{center}
\end{figure}
\begin{figure}[!htb] 
	\begin{center}
		\includegraphics[width=0.48\textwidth]{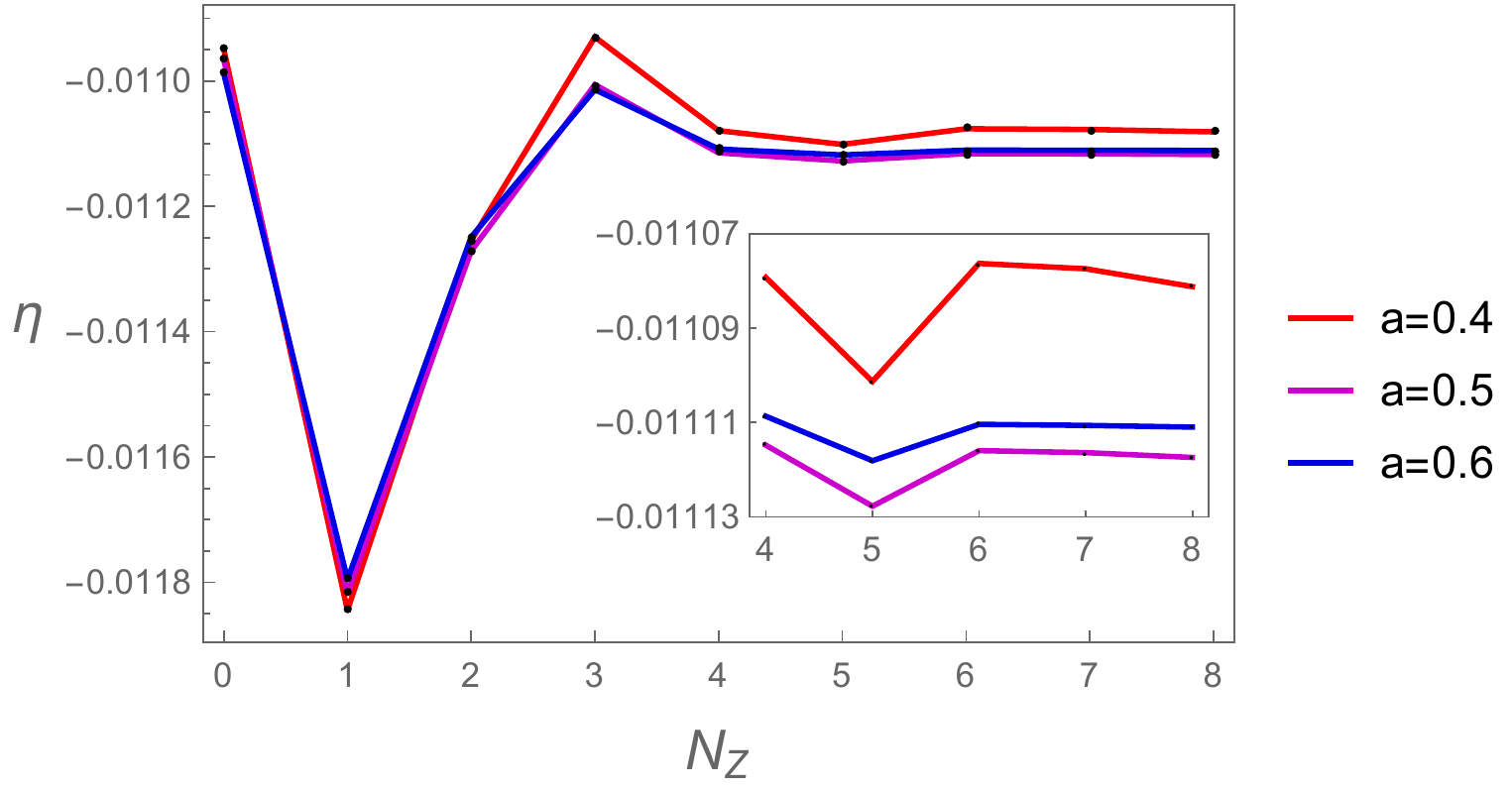}
		\caption{The critical anomalous dimension  $\eta$ in $d=5.9$ as a function of the order of the polynomial truncation.}
		\label{fig:convergence_eta_59D}
	\end{center}
\end{figure}
In addition to $\eta$, it is customary to define the critical exponent
\begin{equation}\label{eq:sigmadef}
 \sigma \equiv \frac{\theta_2}{\theta_1} = \frac{d-2+\eta}{d+2-\eta}\,,
\end{equation}
which is essentially the inverse of the thermodynamical exponent $\delta$
and in fact governs the distribution of the \emph{Yang-Lee edge} as a function of the imaginary magnetic field.

By construction, inserting the classical action into~\Eqref{eq:Wetterich} and expanding in powers of $\lambda_3$ reproduces the
one-loop expressions for $\partial_t \lambda_3$
and $\eta$. These are thus included in the $(N_V,N_Z)=(3,0)$ polynomial truncation,
which comprehends all the classically relevant and marginal couplings, and for which the FP can be easily located at negative values of
 $\lambda_3^{\ 2}$ and $\eta$.
Several strategies are conceivable on how to systematically increase the separate orders
$N_V$ and $N_Z$.
However, as tested in~\cite{Vacca:2015nta} in a similar context, the inclusion of 
interactions in order of their classical dimensionality appears to maximize the convergence
rate. This is to be expected as long as the anomalous dimensions are small enough,
and one can trust the Gau\ss ian notion of relevance of an operator. 
As a consequence, in the following, we will report on the results that descend
from setting $N_V=N_Z+3$ and increasing $N_Z$.

Let us first discuss several properties of the polynomial truncations by remaining close to the
upper critical dimension and setting $d=5.9$.
Already at low orders in $N_Z$, an important issue is the regulator dependence
of universal quantities like $\eta$. The stronger this is, the less accurate
the chosen truncation of theory space~\Eqref{eq:derivative_expansion} is.
In fact, in $d=5.9$, this dependence is rather weak, as shown in Fig.~\ref{fig:eta_vs_a_59D}.
Interestingly, increasing $N_Z$ does not significantly mitigate the regulator dependence of $\eta$.
It is possible to minimize the latter at almost every $N_Z$, 
by locating an $a$-stationary point which appears to converge 
around $a=0.5$ for
larger and larger $N_Z$. 
This numerically agrees with the value of $a$
that minimizes the violation of the second scaling relation,
as is visible in Fig.~\ref{fig:deltalambda2_vs_a_59D}.
By increasing $N_Z$ one can make this violation as small as desired, as expected on the
basis of the exactness of~\Eqref{eq:scaling_rel_2}, and explicitly checked in Fig.~\ref{fig:deltalambda2_vs_NZ}.
Indeed, the series of $\eta$ against $N_Z$ is convergent, 
see Fig.~\ref{fig:convergence_eta_59D}. 
The rate of convergence
does depend on $a$, and it does not seem to be maximal at the point that minimizes the sensitivity of $\eta$,
but close to it, as it can be observed by comparing Figs.~\ref{fig:eta_vs_a_59D} 
and~\ref{fig:convergence_eta_59D}.

\begin{figure}[!htb] 
	\begin{center}
		\includegraphics[width=0.48\textwidth]{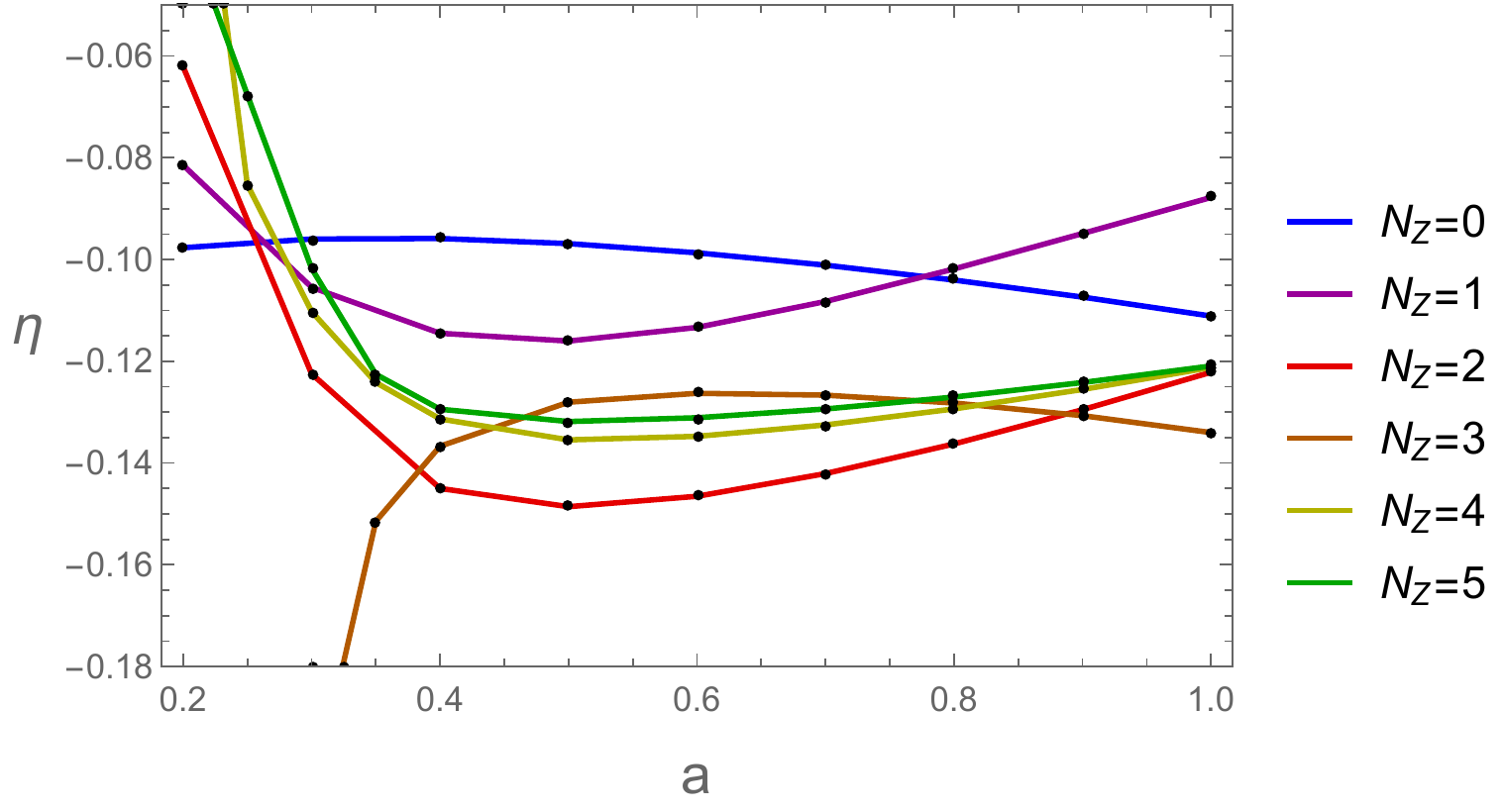}
		\caption{The critical anomalous dimension $\eta$ in $d=5$, as a function of the regulator parameter $a$,
			for several orders of the polynomial truncation.}
		\label{fig:eta_vs_a_5D}
	\end{center}
	\vskip-6mm
\end{figure}

\begin{figure}[!htb] 
	\begin{center}
		\includegraphics[width=0.48\textwidth]{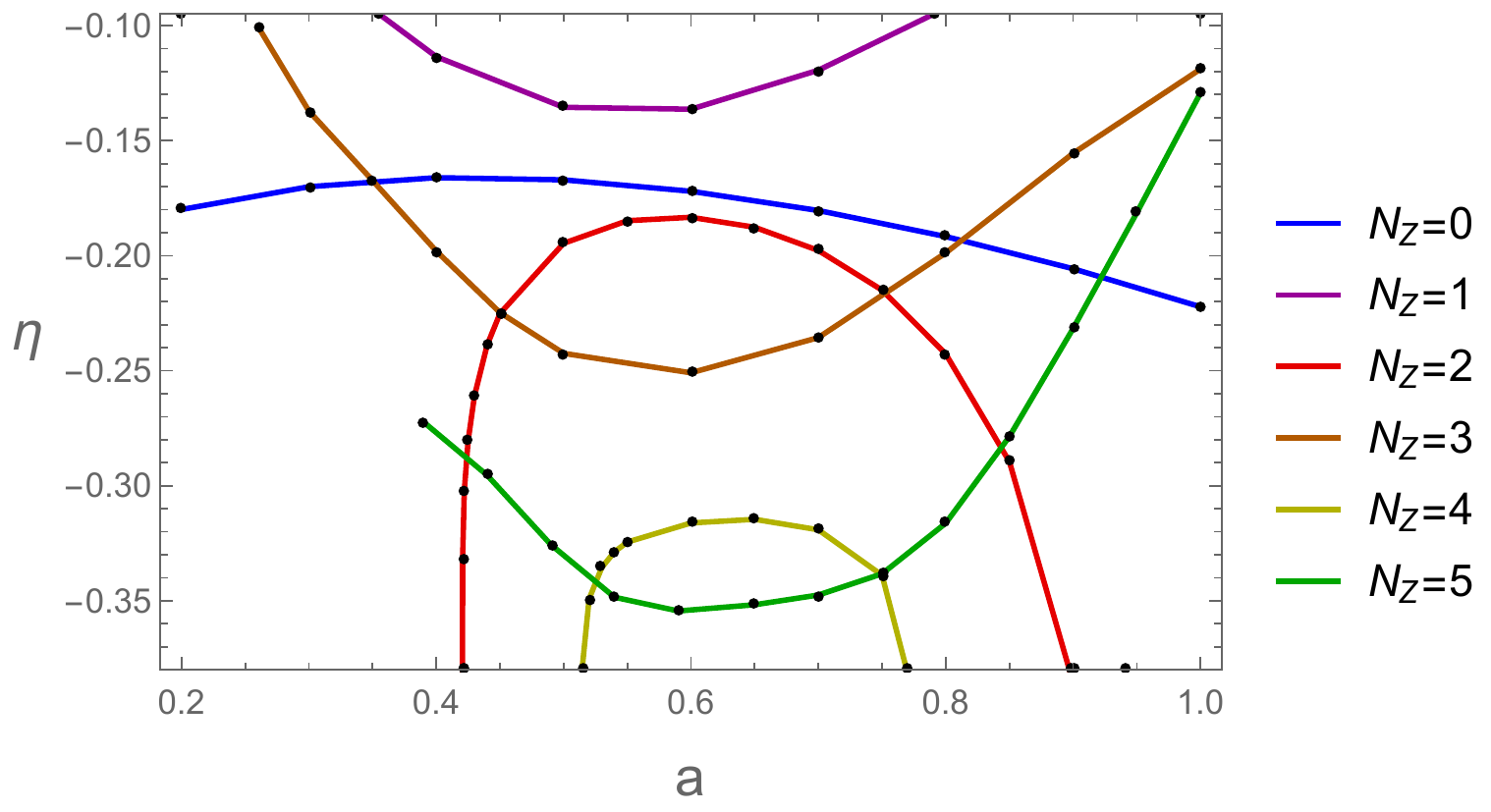}
		\caption{The critical anomalous dimension $\eta$ in $d=4$, as a function of the regulator parameter $a$, 
			for several orders of the polynomial truncation.}
		\label{fig:eta_vs_a_4D}
	\end{center}
\end{figure}

Ensuring the maximum convergence rate becomes crucial for the sake of
lowering $d$. Figures~\ref{fig:eta_vs_a_5D} and~\ref{fig:eta_vs_a_4D} illustrate several important facts.
First, the dependence of $\eta$ on $a$ becomes stronger for lower $d$.
This signals that in lower dimensions the theory becomes
strongly coupled 
\begin{figure}[!htb] 
	\begin{center}
		\includegraphics[width=0.45\textwidth]{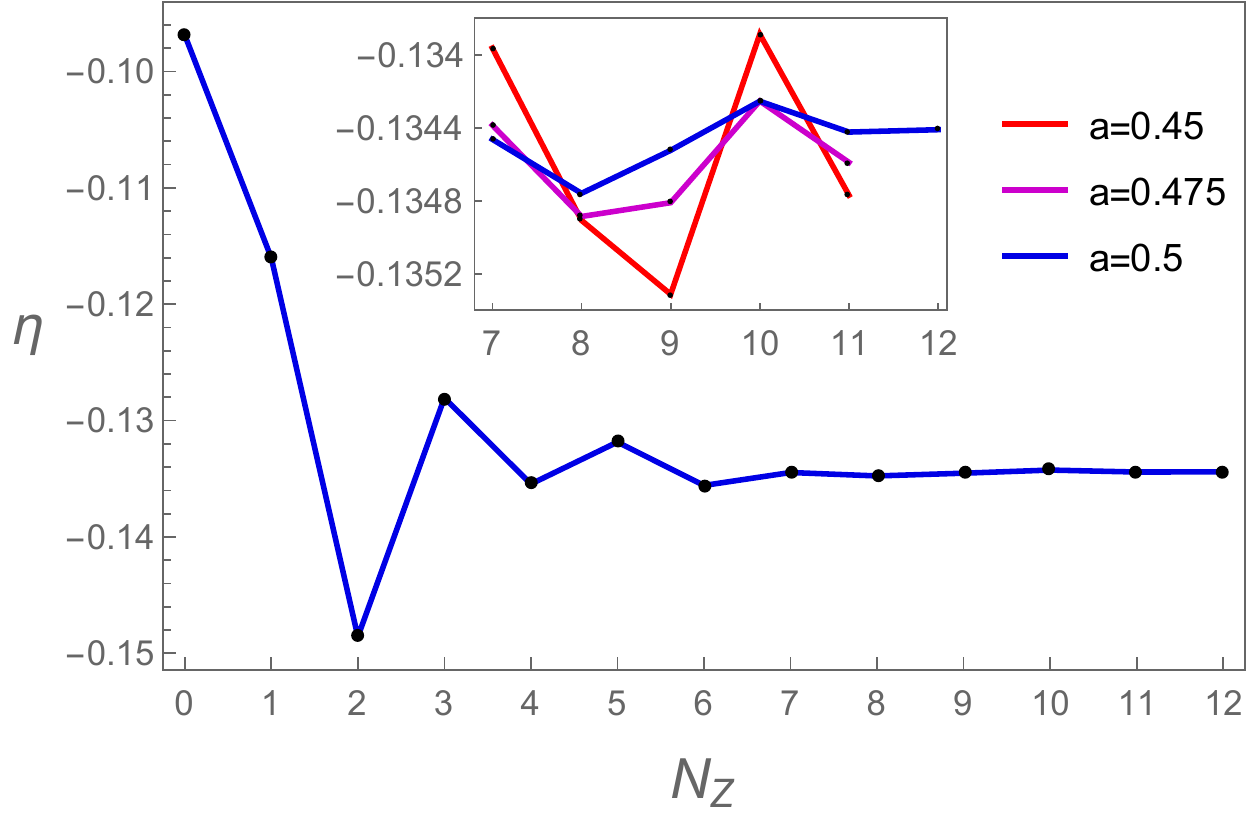}
		\caption{The critical anomalous dimension $\eta$ in $d=5$ as a function of the order of the polynomial truncation.}
		\label{fig:convergence_eta_5D}
	\end{center}
	\vskip-6mm
\end{figure}
\begin{figure}[!htb] 
	\begin{center}
		\includegraphics[width=0.48\textwidth]{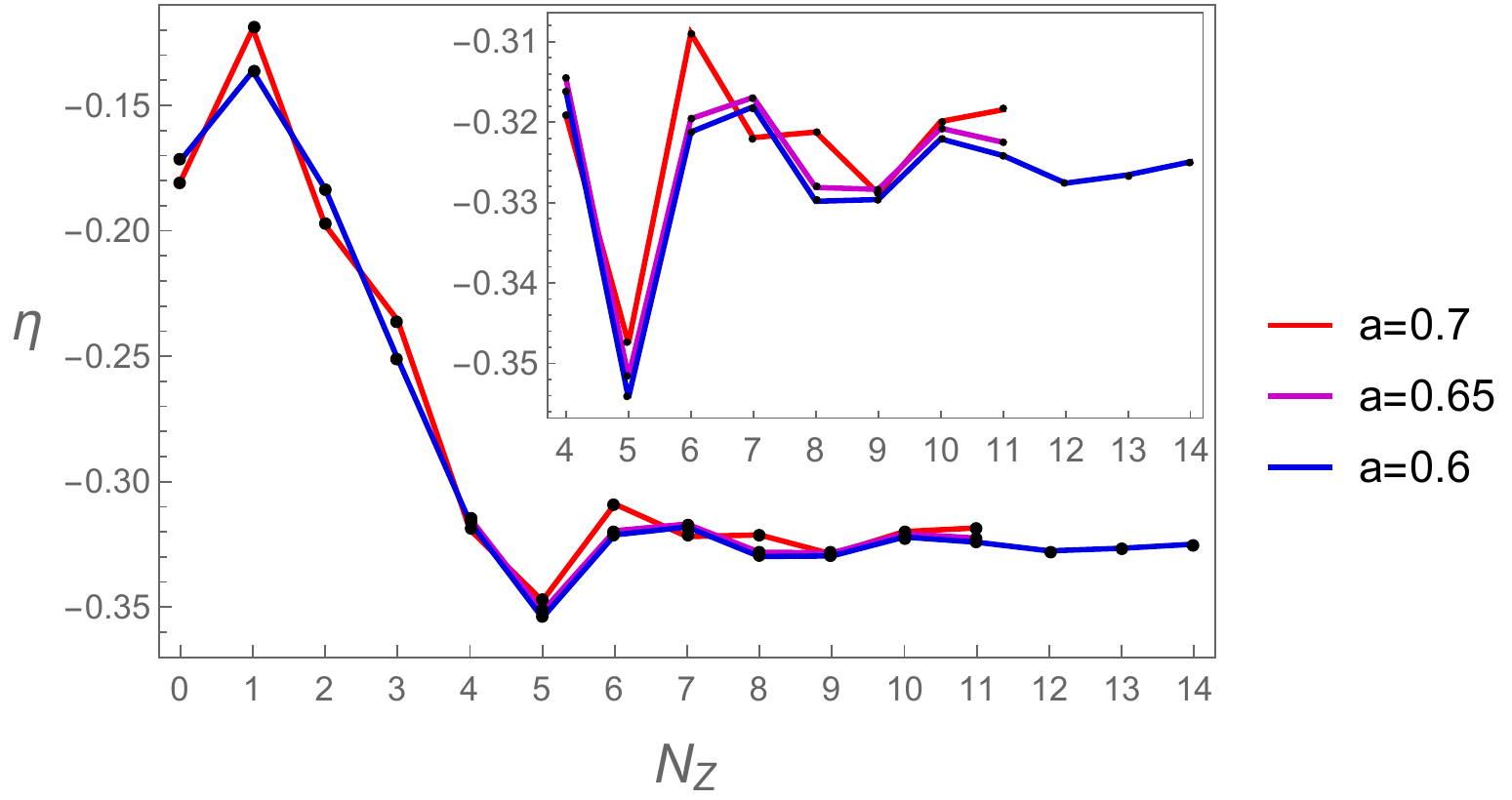}
		\caption{The critical anomalous dimension $\eta$ in $d=4$ as a function of the order of the polynomial truncation.}
		\label{fig:convergence_eta_4D}
	\end{center}
\end{figure}
and the simple parametrization 
of~\Eqref{eq:derivative_expansion}
is less adequate.
Second, the value of $a$ for which the dependence is
minimized moves when $d$ changes. 
This, combined with the technical
problem that following the FP at higher $N_Z$ is difficult in regions
where the $a$-dependence is strong, forces us to continuously relocate
the stationarity point while changing $d$ and $N_Z$.
Let us remark that the minimization of the $a$ dependence of $\eta$
is performed also at orders in $N_Z$ 
much higher than those
displayed in Figs.~\ref{fig:eta_vs_a_5D} and~\ref{fig:eta_vs_a_4D},
yet in a smaller domain.
Third, the overall convergence rate
of the polynomial truncation decreases quickly
for lower $d$,
which can be appreciated by comparing the violation of the second scaling relation
in $d=5.9$ and $d=5$, as shown in the left and right panels of Fig.~\ref{fig:deltalambda2_vs_NZ}, respectively.

Nevertheless, in both $d=5$ and $d=4$ it is possible to see convergence of the 
polynomial truncations, but this requires orders in $N_Z$ which are
much higher than those studied in~\cite{An:2016lni},
where results were obtained up to $N_Z=5$ and $N_V=7$
(though for polynomials centered around a nonzero $\varphi$).
Fig.~\ref{fig:convergence_eta_5D}
shows that in $d=5$ convergence at less than 1\% can be achieved 
already with $N_Z=7$.
Having reached $N_Z=12$ allows us to extract the value
$\eta=-0.1344(1)$ at the optimized point $a=0.5$, within the truncation of~\Eqref{eq:derivative_expansion}.
For $d=4$, at $N_Z=7$ the anomalous dimension still oscillates at the 
level of 4\%, as shown in Fig.~\ref{fig:convergence_eta_4D},
and we need to push the numerics up to $N_Z=14$
to obtain $\eta=-0.325(3)$ at the optimized value $a=0.6$
within the truncation of~\Eqref{eq:derivative_expansion}.

\begin{figure}[!htb] 
	\begin{center}
		% \psfrag{x}{\tiny{$g^2$}}
		% \psfrag{y}{\tiny{$\kappa$}}
		\includegraphics[width=0.45\textwidth]{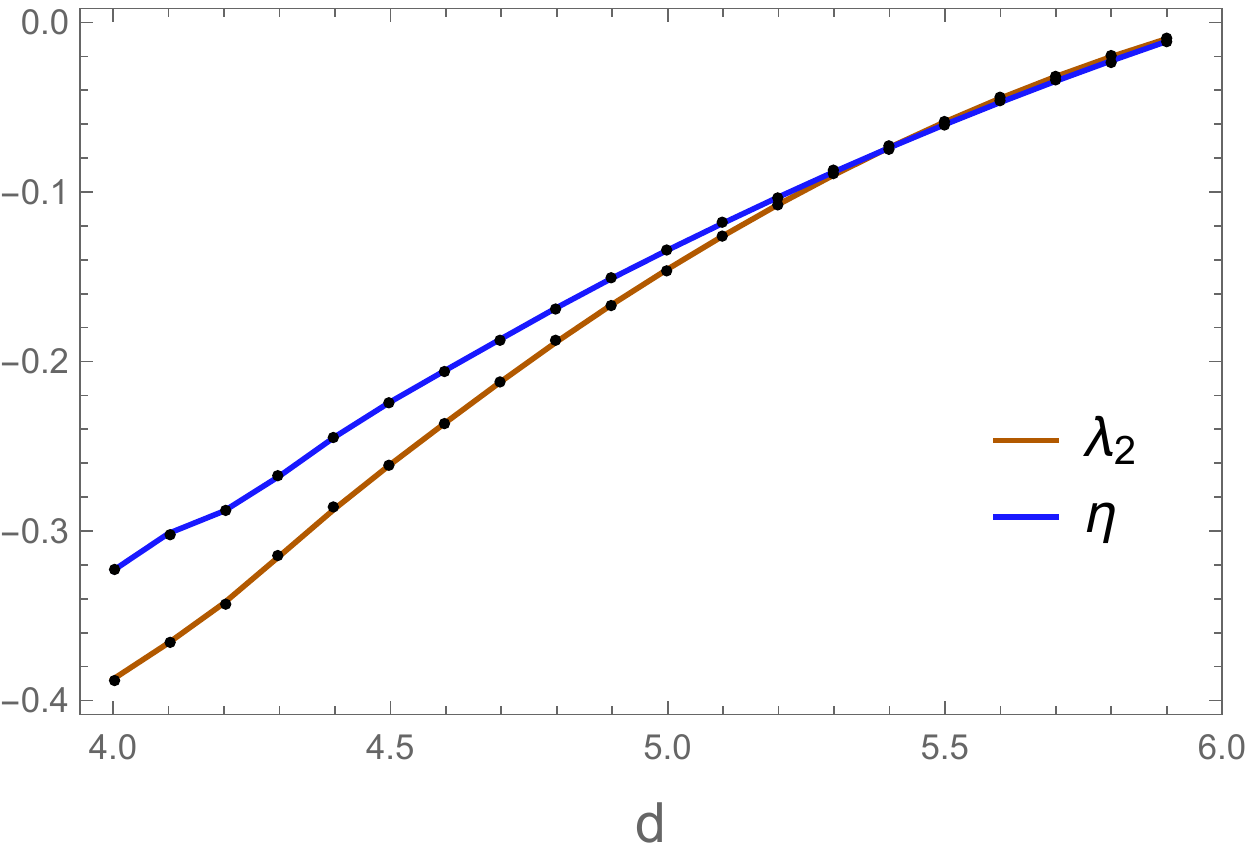}\vskip1mm
		\includegraphics[width=0.45\textwidth]{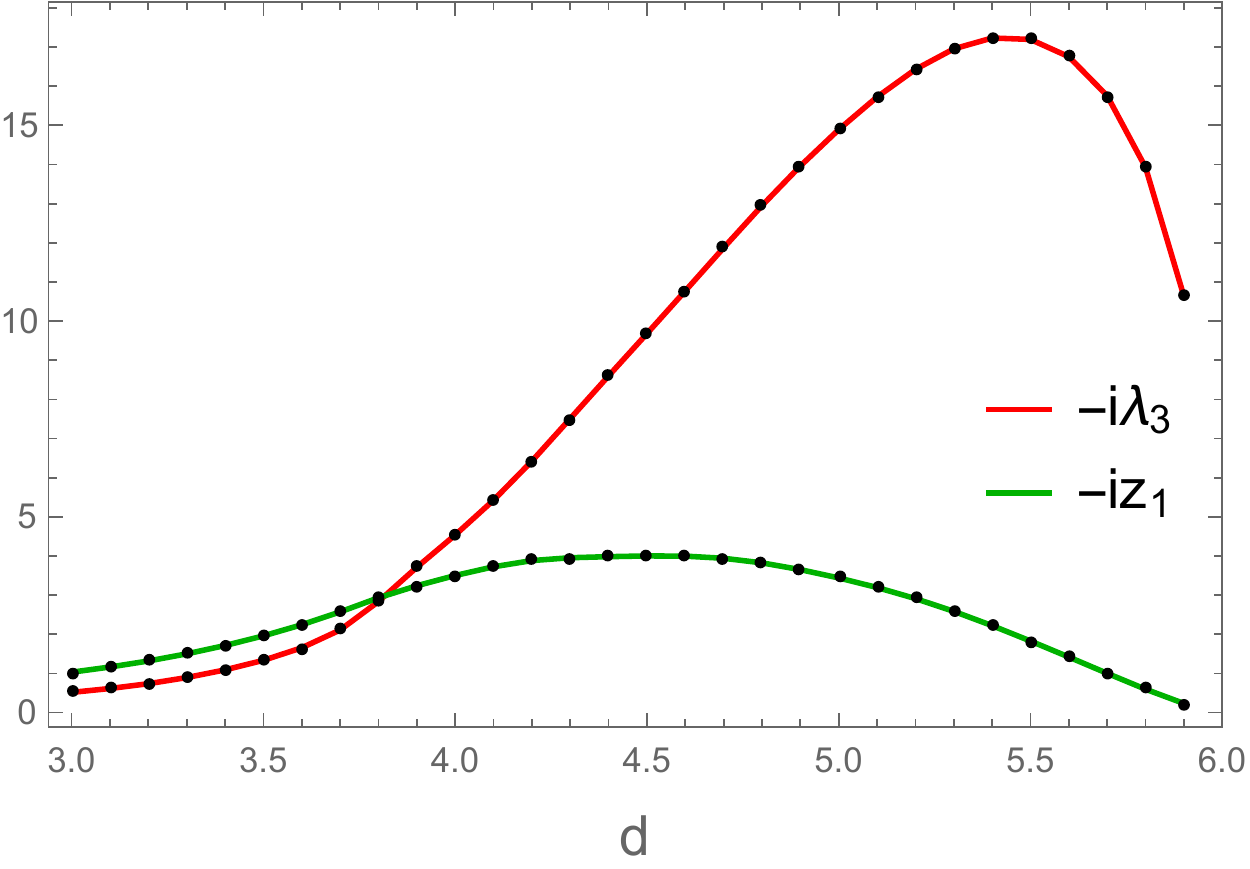}
		\caption{The position of the FP as a function of spacetime dimensions. Upper panel: the mass parameter $\lambda_2$ and the anomalous dimension $\eta$ at order $N_Z=10$, for $a=0.5$.
			Lower panel: the momentum-independent and -dependent cubic couplings  ($\lambda_3$ and $z_1$ respectively) at order $N_Z=7$, for $a=0.7$.}
		\label{fig:FP_vs_d}
	\end{center}
	%\vskip-4mm
\end{figure}

These FPs can be smoothly connected with the perturbative region around $d=6$ by analytically continuing the flow equation in the dimensionality,
such that their identification as nonspurious solutions is
unambiguous. The upper panel of Fig.~\ref{fig:FP_vs_d} illustrates how $\eta$ and the mass parameter $\lambda_2$
 decrease side by side as $d$ is lowered, 
a fact that has no parallel in
the mass-independent scheme usually adopted to construct the $\epsilon$ expansion,
in which $\lambda_2=0$ at the FP.
The lower panel of Fig.~\ref{fig:FP_vs_d} shows that the
cubic couplings $\lambda_3$ and $z_1$ are instead non monotonic.
While in the perturbative region close to $d=6$ the nonderivative 
vertex clearly dominates over the momentum-dependent one,
the two become of the same order of magnitude in $d\leq4$, which indicates that
most likely this applies also to other derivative interactions neglected in
our truncation~\Eqref{eq:derivative_expansion}.
Therefore, it is reasonable to expect that below $d=4$ the present order
of the derivative expansion will be far from accurate.

\begin{figure}[!htb] 
	\begin{center}
		\includegraphics[width=0.45\textwidth]{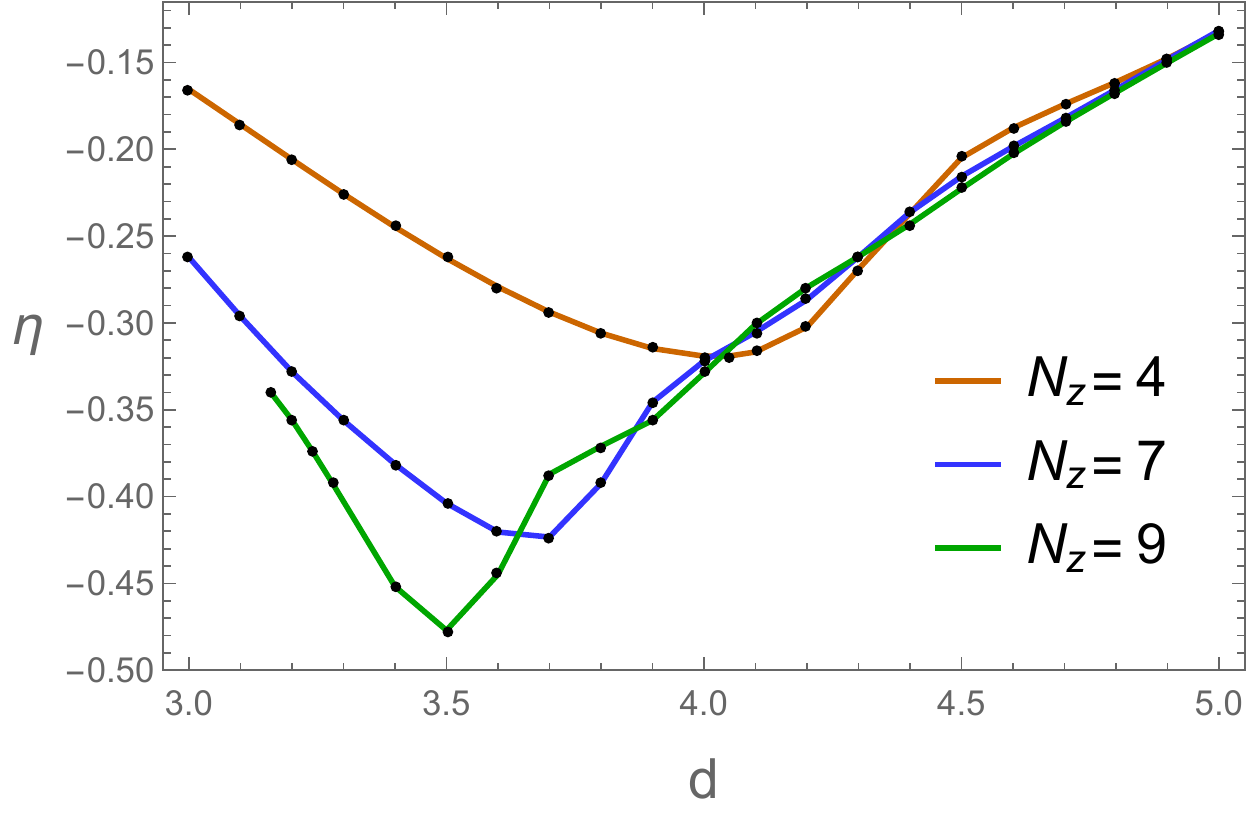}
		\caption{The value of $\eta$ for several orders of the polynomial truncation, as a function of spacetime dimensions $d$. The regulator parameter is fixed at $a=0.7$.}
		\label{fig:eta_vs_d_NZ4_NZ7_NZ9}
	\end{center}
\end{figure}
Unfortunately, we are not able to report on the value of $\eta$ in $d=3$,
because the search for roots of the FP equation becomes extremely 
cumbersome below $d=4$,
 such that in changing $d$, $a$
and $N_Z$ the algorithm is very easily 
attracted to spurious solutions (typically while incrementing $N_Z$) 
or cannot find any FP at all
 (typically   while  lowering $d$).
 To understand the origin of the problem, we can rely only
on low-order truncations, for which probing several $a$ and $d$ values
is much easier.
The results indicate that the main reason for these technical difficulties is 
the slow rate of convergence of the polynomial truncation in lower dimensions,
combined with the stronger $a$-dependence of the FP itself.

Regarding optimization, the minimum-sensitivity value of $a$ grows while $d$ decreases, from $a=0.5$ close to $d=6$ to 
roughly $a=0.7$ in $d=3$. 
Low-order truncations show that for $d<3.8$ the value of $\eta$ undergoes a rapid change across $a\approx0.5$. 
As a consequence, even though the behavior of $\eta$ in $d\leq3.8$ suggests that for $a<0.5$ there might be further stationarity points, we have not been able to systematically probe that region.

Concerning the slow convergence rate, it appears to significantly affect
the $d$-dependence of $\eta$, which is nonmonotonic far from six dimensions. 
As shown in Fig.~\ref{fig:eta_vs_d_NZ4_NZ7_NZ9},
for each order of the polynomial truncation the anomalous dimension reaches a minimum at some $d$ and then it grows in lower dimensions.
The fact that the location of the minimum is pushed toward lower $d$
by increasing $N_Z$, suggests that this nonmonotonicity
is an artifact of polynomial truncations.
Unfortunately, for higher $N_Z$ we are not able to follow the FP
from $d=4$ down to $d=3$, probably because of the combination 
of the increasing complexity of the system of algebraic equations 
and the rapid change of the FP position for $d$ slightly
above the position of the minimum of $\eta$,
as visible in Fig.~\ref{fig:eta_vs_d_NZ4_NZ7_NZ9}.

Finally, from the behavior of the rate of convergence in $4\leq d<6$
we suspect that, even if we could follow the genuine FP down to $d=3$
for higher $N_Z$, to determine $\eta$ with a 
reasonable precision we would still need to probe values of $N_Z$ outside
of our available computing power.
For all these reasons, we believe that the best way to study this FP
in lower dimensions is to abandon polynomial truncations and to use
numerical algorithms such as discretization of field space or
pseudospectral methods~\cite{Borchardt:2015rxa}.
The next question would then be, whether the
$O(\partial^2)$ derivative expansion is able to
give a qualitatively correct description of the Lee-Yang
edge singularity in $d\leq 3$, and if so to which level of accuracy.
So far, we are able to provide an answer only
for $d=5$ and $d=4$, as summarized in Table~\ref{tab:tablenlo}.
We will compare these results to the literature in Sect.~\ref{sec:conclusions}.
\begin{table}[!htb] 
	\begin{tabular}{ l | c  | c |}
		& $\eta$  & $\sigma$\\
		\hline
		${\rm LY}_5$ & $-$0.1344(1) & 0.40166(2) \\
		${\rm LY}_4$ & $-$0.325(3) & 0.2648(6)\\
		\hline
	\end{tabular}
	\caption{Summary of the estimates coming from the $O(\partial^2)$ of the derivative expansion.}
	\label{tab:tablenlo}
\end{table}
%
%

%%%%%%%%%%%%%%%%%%%%%%%%%%%%%%%%%%%%%%%%
\section{Local potentials and nonunitary multicriticality}
\label{sec:imaginary_approximation}
%%%%%%%%%%%%%%%%%%%%%%%%%%%%%%%%%%%%%%%%

In this section we describe an approximation that is inspired by the original work by Fisher \cite{Fisher:1978pf},
which is probably the simplest description of the Lee-Yang model in terms of a local potential.
Notably, it will allow us to extend our investigation to two and three dimensions.
We concentrate our attention to local potential truncations of \eqref{eq:derivative_expansion} for which $Z_k(\varphi)={\rm const.}$
The flow of the dimensionless renormalized potential becomes considerably simple if the cutoff function \eqref{eq:cutoff_p2} is used at $a=1$
\begin{equation}\label{eq:vdotsimple}
\begin{split}
 \partial_t v(\varphi) &=- dv(\varphi) + \frac{d-2+\eta}{2}\varphi v'(\varphi)  + c_d \frac{1-\frac{\eta}{d+2}}{1+v''(\varphi)}\,,
\end{split}
\end{equation}
with $c_d^{-1}=(4\pi)^{d/2}\Gamma(1+d/2)$. Having fixed $a$, from now on we will not be able to study the sensitivity of the results on the cutoff.
It is easy to see from the right hand side of \eqref{eq:vdotsimple} that the stationarity condition $\partial_t v(\varphi)=0$ can be used to generalize FPs to solutions of a nonlinear ordinary differential equations \cite{Hellwig:2015woa}.

For a general real cutoff, and in particular for our choice \eqref{eq:vdotsimple},
the Wetterich equation \eqref{eq:Wetterich} is invariant under a generalization of $\mathbb{Z}_2$ parity
known as ${\cal PT}$ symmetry. In terms of the dimensionless renormalized potential we define its action as
\begin{equation}
\begin{split}
 {\cal PT}: v(\varphi) &\to v(-\varphi)^* \,,
\end{split}
\end{equation}
where the star indicates complex conjugation.
It is natural to separate the real and imaginary parts of the potential $v(\varphi)\equiv u(\varphi)+i h(\varphi)$,
\begin{figure}[!htb] 
	\begin{center}
		% \psfrag{x}{\tiny{$g^2$}}
		% \psfrag{y}{\tiny{$\kappa$}}
		\includegraphics[width=0.45\textwidth]{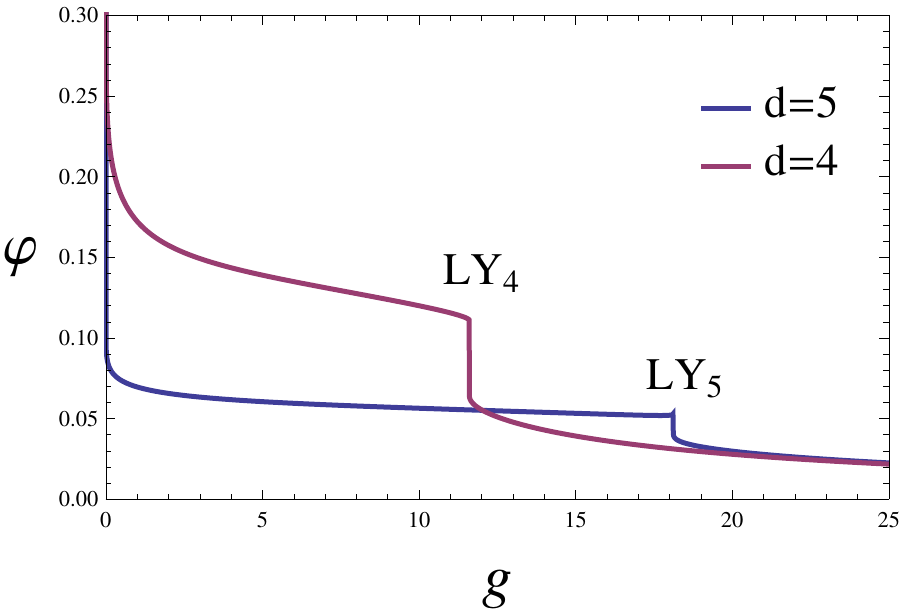}
		\caption{Terminal values of the integration of the FP condition in $d=5$ and $d=4$ dimensions for varying boundary condition $g=-i\lambda_3$.
			The singularities corresponding to the Lee-Yang model are highlighted for both dimensionalities by insets.}
		\label{fig:spike_plot_d4_and_d5}
	\end{center}
\end{figure}
and use this symmetry to argue that for invariant stationary solutions $v(\varphi) = v(-\varphi)^*$ the real part of the potential is an even function, while the imaginary part is odd:
\begin{eqnarray}\label{boundary_conditions}
 u(-\varphi) = u(\varphi) & \quad {\rm and } \quad &  h(-\varphi) = -h(\varphi) \,.
\end{eqnarray}
It is possible to decompose the flow \eqref{eq:vdotsimple} into a flow for real and imaginary parts as $\partial_t v(\varphi)=\partial_t u(\varphi)+i\partial_t h(\varphi)$.
The even potential $u(\varphi)$ could be thought to contain the information of an Ising-like model,
while the odd $h(\varphi)$ can be interpreted as a functional generalization of an imaginary magnetic field.

Fisher showed  that the critical exponents of the Yang-Lee edge singularity in the thermodynamical limit
can be related to the scaling exponents of the Lee-Yang model by opportunely tuning the imaginary magnetic field to criticality \cite{Fisher:1978pf}.
We shall mimic this setting by truncating the effective potential to its imaginary part, thus effectively setting the real part to zero: $u(\varphi)=0$.
We are left with the flow of a single function
\begin{equation}\label{eq:hdotsimple}
\begin{split}
 \left.\partial_t h(\varphi)\right|_{u=0} &= -dh +\frac{d-2+\eta}{2}\varphi h'  - c_d h'' \frac{1-\frac{\eta}{d+2}}{1+h^{\prime\prime2}}\,,
\end{split}
\end{equation}
the stationary solutions of which can be studied with the boundary condition $h(0)=0$ coming from \eqref{boundary_conditions}. In this approximation there is only one parameter left to tune to criticality,
which we choose to be the third derivative
$$g\equiv -i\lambda_3=-i v'''(0)=h'''(0)\,.$$
A direct comparison of \eqref{eq:hdotsimple} with the flow of the ${\cal N}=1$ superpotential $w(\varphi)$ in $(d+1)$ dimensions in the same scheme~\cite{Synatschke:2009nm,Hellwig:2015woa} shows the structural similarities between the two problems,
which might be related to some generalization of the arguments of~\cite{Parisi:1979ka}.

The problem is now apt to the application of the standard shooting methods described in \cite{Hasenfratz:1985dm,Morris:1994ki,Codello:2012sc,Hellwig:2015woa} to find the critical values for the parameter $g$.
\begin{figure}[!htb] 
	\begin{center}
		% \psfrag{x}{\tiny{$g^2$}}
		% \psfrag{y}{\tiny{$\kappa$}}
		\includegraphics[width=0.45\textwidth]{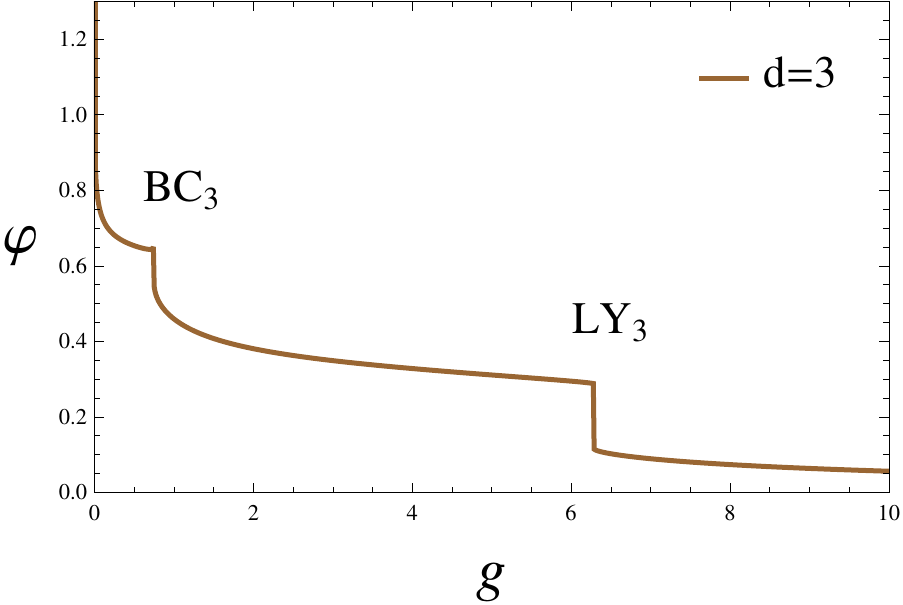}
		\caption{Terminal values of the integration of the FP condition in $d=3$ dimensions for varying boundary condition $g=-i\lambda_3$.
			The singularities corresponding to the Lee-Yang and the Blume-Capel models are highlighted by insets.}
		\label{fig:spike_plot_d3}
	\end{center}
\end{figure}
 In brief, the method consists in numerically integrating the second-order ordinary differential equation
obtained by setting
the right hand side of  \eqref{eq:hdotsimple} equal to zero, with varying boundary conditions parametrized by $g$. 
For all values of $g$, the integration terminates in a singularity for (the third derivative of) $h(\varphi)$,
which occurs at a finite value of $\varphi$. The plot of this terminal value of the field as a function of $g$ displays spikes and 
flexes in proximity of the critical values, 
indicating the existence of solutions that can be extended to all values of the field. We refer to \cite{Hellwig:2015woa} and references therein for more details on the approach, including its relation to the Ginzburg-Landau theory.

The plot of the terminal value for the cases $d=5$ and $d=4$ are given in Fig.\ \ref{fig:spike_plot_d4_and_d5} as a function of $g$.
We highlighted the location of the singularities associated to the Lee-Yang model for both dimensionalities.
The values for $g$ are comparable to those shown in Fig.\ \ref{fig:FP_vs_d}, but are not quantitatively similar.\footnote{A less crude approximation involves truncating the even part of the potential to a mass term
$u(\varphi)=\frac{1}{2}\lambda_2\varphi^2$, and tuning $\lambda_2$ to criticality using $\partial_t u=0$, i.e. solving $\partial_t\lambda_2=0$ for $\lambda_2$ as a function of $g$ too. This procedure gives a better agreement for nonuniversal quantities such as $g$, but it can be shown that it does not change the critical exponents $\eta$ and $\sigma$. Since the physical content of the model is in its critical exponents, we decided to leave out this additional complication.} The singularity corresponding to $g=0$ is the Gau{\ss}ian model.

The plot for the case $d=3$ is given in Fig.\ \ref{fig:spike_plot_d3}. The singularity corresponding to the Lee-Yang model can be followed easily from the higher dimensionalities, and it is still evident on the right of the plot. However, in $d=3$ a new critical theory is present on the left of
 the Lee-Yang model. We believe that the new theory corresponds to the quintic $i\varphi^5$ model, the upper critical dimension of which is $d=10/3$ and therefore can be non-Gau{\ss}ian in three dimensions. 
To test this hypothesis we explicitly checked that its singularity spawns from the Gau{\ss}ian solution while crossing its upper critical dimension $d=10/3$, thus agreeing with Helmholtz theorem and the standard arguments of mean field. The quintic model is in the same universality class of the Blume-Capel model \cite{vonGehlen:1994rp,Belavin:2003pu}, which has a tricritical phase at an imaginary magnetic field.\footnote{The original Blume-Capel model is the one of a spin chain, thus it should be clear that this universality class refers to the same model on a two-dimensional grid evolving in time. While with our nomenclature convention the tricritical Ising universality class could also be referred to as Blume-Capel model, we decided to use it for the lesser-studied theory.} 
Following a standard practice, we took the liberty of naming it simply the Blume-Capel model, even though it is implied that by model we mean universality class.
For the most part, the Blume-Capel model has been studied in two spacetime dimensions, corresponding to one-dimensional spin chains, but the model is critical (albeit nonunitary) in three dimensions too, and thus furnishes a rather little-studied example of a three-dimensional critical theory.

The latter discussion on the quintic model can be easily generalized to all models governed by odd powers of the field $i\varphi^{2n+1}$ for $n\in \mathbb{N}_0$,
which include the Lee-Yang and the Blume-Capel cases for $n=1$ and $n=2$, respectively. Their upper critical dimensions are
\begin{equation}\label{eq:uppercritical}
\begin{split}
 d_n &= 2+\frac{4}{2n-1} \,.
\end{split}
\end{equation}
In the limit $n\to \infty$, the critical dimensionalities approach $d_n\to 2$, that is, the special case in which the field $\varphi$ becomes canonically dimensionless.
We conjecture that this sequence of critical models interpolates with the sequence of minimal conformal theories ${\cal M}(2,2n+3)$ in the two-dimensional limit \cite{Belavin:2003pu}.
This statement is backed by both the structure of the operator product expansion, as well as the fact that ${\cal M}(2,5)$ and ${\cal M}(2,7)$ are known descriptions of the Lee-Yang and Blume-Capel models respectively \cite{Cardy:1985yy,vonGehlen:1994rp}.

Following the emergence of these models in dimensions lower than three is, however, more complicated than observing the previous cases: the cutoff \eqref{eq:cutoff_p2} has poorer convergence properties than usual when approaching two dimensions in \eqref{eq:hdotsimple}. The convergence can be improved by changing the power of the momentum $p^2$ of \eqref{eq:cutoff_p2} as described for the supersymmetric flows in \cite{Hellwig:2015woa}. All the steps described above and leading to the flow \eqref{eq:hdotsimple} for the imaginary part can still be performed, but are much more complicated and lead to a highly nonlinear equation, which considerably complicates the numerical analysis.
Nevertheless, we explicitly checked that the first few $i\varphi^{2n+1}$ models spawn from the Gau{\ss}ian solution when lowering $d$ to values smaller than their critical dimensions, in a guise similar to what we observed for the Blume-Capel model.

In analogy with the relation between the Ising and the Lee-Yang universality classes, it can be speculated that the Blume-Capel class might be relevant in elucidating the structure of the singularities in the complex activity plane of the tricritical Ising model \cite{Mossa:2007fx}.  
We hope to report more on the Blume-Capel universality class and on the topic of multicriticality at imaginary values of the magnetic field in the future.
All the results of this section for $d=5,4,3,2$ are summarized in Tab.\ \ref{tab:tablespikes}.
\begin{table}[!htb] 
	\begin{tabular}{ l | c  | c |}
		& $\eta$  & $\sigma$\\
		\hline
		${\rm LY}_5$ & $-$0.176 & 0.394 \\
		${\rm LY}_4$ & $-$0.426 & 0.245\\
		${\rm LY}_3$ & $-$0.667 & 0.0588 \\
		${\rm LY}_2$ & $-$0.958 & $-$0.193 \\
		\hline
		${\rm BC}_3$ & $-$0.0093 & 0.198 \\
		\hline
	\end{tabular}
	\caption{Summary of the estimates coming from the Fisher-inspired approximation \eqref{eq:hdotsimple}.
		Models are labeled by their initials while the subscripts refer to their dimensionality.}
	\label{tab:tablespikes}
\end{table}
\begin{table*}%[!htb] 
	%	\centering
	\begin{tabular}{ l | c | c | c | c | c | c | c |}
		& this work  &\cite{An:2016lni} &\cite{Gracey:2015tta} &\cite{Gliozzi:2014jsa}	&\cite{Butera:2012tq}	&\cite{Lai:1995}	&\cite{Hsu:2005}\\
		\hline
		${\rm LY}_5$ & 0.40166(2) & 0.4033(12)	& 0.3981 & 0.4105(5)	& 0.401(9)	& 0.402(5)	& 0.40(2)	\\
		${\rm LY}_4$ & 0.2648(6)  & 0.2667(32)	& 0.2584 & 0.2685(1)	& 0.258(5)	& 0.2648(15)	& 0.261(12)	\\
		\hline
	\end{tabular}
	\caption{Estimates of the critical exponent $\sigma$ at the Yang-Lee edge singularity in $d=5$ and $d=4$,
		from this work as well as
		from an exploratory functional RG study~\cite{An:2016lni},
		the $\epsilon$ expansion at four loops~\cite{Gracey:2015tta},
		the conformal bootstrap~\cite{Gliozzi:2014jsa},
		strong coupling expansions~\cite{Butera:2012tq}, 
		Monte Carlo simulations of hard-core fluids~\cite{Lai:1995}, and
		of lattice animals and trees~\cite{Hsu:2005}. The error of our estimates is to be understood as purely numerical,
		and it does not include an estimate of the systematic truncation effects.}
	\label{tab:sigma}
\end{table*}
%

%%%%%%%%%%%%%%%%%%%%%%%%%%%%%%%%%%%%%%%%
\section{Conclusions}
\label{sec:conclusions}
%%%%%%%%%%%%%%%%%%%%%%%%%%%%%%%%%%%%%%%%

We gave functional renormalization group estimates for the critical properties of the Lee-Yang model in less than six dimensions.
Using a $O(\partial^2)$ truncation of the space of all operators, and carefully optimizing the numerics, we provided converging estimates to the anomalous dimension and the exponent $\sigma$ in five and four dimensions as summarized in Tab.\ \ref{tab:tablenlo}.
It is important to stress that we do not estimate the truncation error, only the error related to the numerics.
Our results are compared to some of the estimates produced by other methods in Tab.~\ref{tab:sigma}.
These include another FRG study using a different scheme as well as different lower-order polynomial truncations~\cite{An:2016lni},
the four-loop $\epsilon$ expansion with a Pad\'e approximant constrained by the
exact results in $d=1$ and $d=2$~\cite{Gracey:2015tta},
the conformal bootstrap~\cite{Gliozzi:2014jsa},
the strong coupling expansion~\cite{Butera:2012tq},
and simulations of fluid models with repulsive-core interactions~\cite{Lai:1995} and of lattice animals and trees~\cite{Hsu:2005}.
Thanks to the strong constraints that the scaling relations enforce on it, all the relevant properties of the Lee-Yang model are determined by  $\eta$.

Unfortunately, we could not provide equally reliable estimates in three and two dimensions. Based on our analysis, we argued that in this case the problem is both technically challenging, because much bigger truncations are needed to ensure convergence, but also that the simple next-to-leading order of the derivative expansion might be insufficient.
We believe that a reliable solution to this problem might lead to a deeper understanding of the limitations of the derivative expansion
and of how to appropriately choose truncations of the effective action of which the scope is the computation of critical exponents.

We then showed that there exists a simple, yet very useful approximation of the effective potential which allows us to qualitatively explore
the three- and two-dimensional cases.
This is well suited to describe the appearance of further nonunitary critical solutions, which we argued to represent the sequence of models $i\varphi^{2n+1}$,
and which we conjectured to interpolate with a well known family of conformal theories in two dimensions.
While the first element of this sequence is the Lee-Yang model,
the second is a tricritical model which we conjectured to describe the critical Blume-Capel model on a spin-lattice at imaginary values of the magnetic field.
This model happens to have an upper critical dimension higher than three, and thus shows a non-Gau\ss ian behavior in three dimensions,
giving a less known example of a three-dimensional critical theory of a single scalar field.

%%%%%%%%%%%%%%%%%%%%%%%%%%%%%%%%%%%%%%%%
\section*{Note added}
%%%%%%%%%%%%%%%%%%%%%%%%%%%%%%%%%%%%%%%%

After the publication of this work we became aware of a work discussing the Landau-Ginzburg description of some of the nonunitary two-dimensional minimal models \cite{tesi-amoruso},
the conclusions of which do not agree with the conjecture made in this paper that the Blume-Capel (quintic) universality class in $d>2$ interpolates with the minimal model ${\cal M}(2,7)$ in $d=2$.
In \cite{tesi-amoruso} it is conjectured that the minimal model
${\cal M}(2,9)$ admits a $i\varphi^5$ Landau-Ginzburg description. This agrees with the fact that the quintic theory is expected to correspond to a theory that
has four relevant \emph{and} primary operators in its Kac table (here we include the identity and exclude the operator $\varphi^4$ because it is not a primary,
but rather a descendant due to the equations of motion $ \partial^2\varphi \sim \varphi^4$).

A variant of the conjecture stated in this paper, which also accomodates the recent findings, would be to argue that the odd multicritical models $i\varphi^{2n+1}$
could be the correct Landau-Ginzburg descriptions of the sequence ${\cal M}(2,4n+1)$ of CFT minimal models.
This second conjecture agrees with the Lee-Yang case which occurs at $n=1$, but also agrees with the general discussion of \cite{tesi-amoruso} and specifically with the $n=2$ case,
and resonates with the general expectation that the $i\varphi^{2n+1}$ model should correspond to a model that has $2n$ relevant and primary operators in its Kac table.

The conjecture enunciated here is part of the more general quest for a Landau-Ginzburg classification of all two-dimensional CFTs
and especially of all minimal models (see \cite{tesi-amoruso} and references therein).
We believe that such a classification is an objective of paramount importance
for the development of QFT and statistical mechanics, and we hope to give a contribution to this topic in the future.
% \newline

%%%%%%%%%%%%%%%%%%%%%%%%%%%%%%%%%%%%%%%%
\section*{Acknowledgments}
%%%%%%%%%%%%%%%%%%%%%%%%%%%%%%%%%%%%%%%%

LZ acknowledges support by the DFG under grant GRK 1523/2.
OZ acknowledges support by the DFG under grants Gi 328/7-1 and Gi 328/6-2 (FOR 723),
and hospitality of G.\ P.\ Vacca and INFN Bologna during the completion of this work.
OZ is grateful to G.\ Mussardo for an important clarification during the development of this work.

\end{document}